\documentclass[3p,times,review]{elsarticle}
\usepackage{ecrc}
\usepackage{float}
\floatstyle{ruled}
\newfloat{algorithm}{tbp}{loa}
\floatname{algorithm}{Algorithm}

\volume{00}

\firstpage{1}

\journalname{Digital Signal Processing}

\runauth{}

\jid{procs}

\jnltitlelogo{digital Signal Processing}

\CopyrightLine{2011}{Published by Elsevier Ltd.}

\usepackage{amssymb}

\usepackage[figuresright]{rotating}

\begin{document}
\begin{frontmatter}
%
\title{The Recursive Gauss-Newton Filter}
%
%
%

\author{Roaldje Nadjiasngar}
\address{Roaldje.Nadjiasngar@uct.ac.za}
\author{ Michael Inggs}

\address{Michael.Inggs@uct.ac.za}

\begin{abstract}
This paper presents a compact, recursive, non-linear, filter, derived from the Gauss-Newton (GNF), which is an algorithm that is based on weighted least squares and the Newton method of local linearisation. The recursive form (RGNF), which is then adapted to the Levenberg-Maquardt method is applicable to linear / nonlinear of process state models, coupled with the  linear / nonlinear observation schemes.  Simulation studies have demonstrated the robustness of the RGNF, and a large reduction in the amount of computational memory required, identified in the past as a major limitation on the use of the GNF. 
\end{abstract}



\end{frontmatter}

%

\section{Introduction}

The minimum variance algorithm has been used to estimate parameters from batches of observations, accumulated over a defined period of time. The most popular version of the minimum variance methods is the weighted  least squares, which are at the heart of adaptive filtering \cite{Yunmin:99} \cite{Thomas:00}. The recursive least squares (RLS) methods are efficient 
versions  of the least squares approach, and are applicable to estimation of future states from scalar input data  streams. However, recent studies \cite{malik:04} have seen the development of state space recursive least squares (SSRLS) methods that show robustness in the estimation of linear state space models. For the estimation of non-linear state space models, a non-recursive filter called the Gauss-Newton filter (GNF)   was developed and has
been successfully used in many applications \cite{Morrison:newbook} \cite{Morrison:2007a} .
The GNF algorithm is a combination of the Newton method of local linearization and
the least squares-like version of the minimum variance method\cite{Morrison:newbook}. It is used to estimate process states that are governed by non-linear, autonomous,  differential equations, coupled with linear or non-linear observation schemes. The GNF algorithm, although robust, requires significant processing 
power, i.e.  the amount of memory required. To improve the computational efficiency of the GNF, studies of the use of  Field Programmable Gate  Arrays (FPGA) and other co-processor technology have been made \cite{jpdacon:11,Milburn:2010a}. Memory requirements were identified in these studies as being the major stumbling block in implementations on both on FPGA (low power and parallelism) and coprocessor (ease of use) technology. This paper  obtains a recursive form of the GNF with zero memory. We then adapt the recursive filter to the Levenberg-Maquardt method, renown for its robustness \cite{Mquardt,LMA,ScienceDLMapplications,ScienceDLMapplications2,ScienceDLMapplications3}, widely used in non linear curve fitting problems and neural networks algorithms.  The contribution of this paper is the derivation of a compact recursive form of the GNF  that is applicable to four major scenarios:
\begin{description}
\item Case 1 : linear process dynamic and linear observation scheme. 
\item Case 2 : linear process dynamic and non-linear observation scheme. 
\item Case 3 : non-linear process dynamic and linear observation scheme. 
\item Case 4 : non-linear process dynamic and non linear observation scheme. 
\end{description}
The paper begins with an exposition of a state space model based on non-linear, differential equations. This is followed, in Section\ \ref{sec:recurse}, by the derivation of a recursive GNF. In Section \ref{sec:Maquard} we describe the adaptation of the recursive equations of the filter  to the Levenberg-Maquardt method. A complete filter algorithm is presented. In Section\ \ref{sec:ssm}, the state space situations to which we can apply this new recursive form are demonstrated, with a look at stability.  We then demonstrate the power of the new recursive GNF in an application to range and bearing only tracking of a manoeuvring target (Section\ \ref{sec:demo}), before concluding with a summary of results achieved.

\section{State space model based on non-linear differential equations\label{sec:stateSpace}}

Consider the following autonomous, non-linear differential equation (DE) governing 
the process state:

 \begin{equation}
DX(t)=F(X(t))\end{equation}

in which $F$ is a non linear vector function of the state vector $X$ describing a process, such as the position of a target in space.
We assume the observation scheme of the process is a non-linear function
of the process state  with expression :

\begin{equation}
Y(t)=G(X(t))+v(t)\end{equation}

where $G$ is a non-linear function of $X$ and $v(t)$ is a random Gaussian vector. 
The goal is to estimate the process state from the given non-linear  state models. 
For linear differential equations (DEs),  the state transition matrix  could be easily obtained. This, however, is not the case with non-linear DEs. Nevertheless, there is a 
procedure, based on local linearization, that enables us to get around this obstacle, which we will now present.

\subsection{The method of local linearisation}
The solution of the DE gives rise to infinitely many trajectories that are dependent 
on the initial condition. However there will be one trajectory whose state vector
the filter  will attempt to  identify from the observations. We assume 
that there is a known nominal trajectory with state vector $\bar{X}(t)$  that has the following properties:

 \begin{itemize}
	\item $\bar{X}(t)$ satisfies the same DE as  $X(t)$ 
	\item $\bar{X}(t)$ is close to $X(t)$
\end{itemize}

The above-mentioned properties result in the following expression:

\begin{equation}
X(t)=\bar{X}(t)+\delta X(t) 
\end{equation}

where $\delta X(t)$ is a vector of time-dependent functions that are small 
in relation to the corresponding elements of either $\bar{X}(t)$ or $X(t)$ . 
The vector $\delta X(t)$ is called the {\em perturbation vector} and is governed by the 
following DE  (the derivation is shown in Appendix A):

 \begin{equation}
D(\delta X(t))=A(\bar{X}(t))\delta X(t)\end{equation}

where $A(\bar{X}(t))$ is called a sensitivity matrix  defined as follows:

\begin{equation}
A(\bar{X}(t))=\left.\frac{\partial F(X(t))}{\partial(X(t))}\right|_{\bar{X}(t)}
\end{equation}. 

Equation is thus a linear DE, with a time varying coefficient and has a 
the following transition equation:

\begin{equation}
\delta X(t+\zeta)=\Phi (t_{n}+\zeta,t_{n},\bar{X})\delta X(t)
\end{equation}

in which  $\Phi (t_{n}+\zeta,t_{n},\bar{X})$ is the transition matrix from time 
$t_{n}$ to $t_{n}+\zeta$  (increment $\zeta$). The transition matrix is 
governed by the following DE:

\begin{equation}
\frac{\partial}{\partial\zeta}\Phi(t_{n+\zeta},t_{n},\bar{X})=
A(\bar{X}(t_{n}+\zeta))\Phi(t_{n+\zeta},t_{n},\bar{X})
\end{equation}

\begin{equation}
\Phi(t_{n},t_{n},\bar{X})=I 
\end{equation}

The transition matrix is a function of $\bar{X}(t)$  and can be evaluated by 
numerical integration and in order to fill the values of $A(\bar{X}(t_{n}+\zeta))$,
 $\bar{X}(t)$ has to be integrated numerically. We will soon present a recursive
 algorithm that will avoid the computation of the transition matrix. We have 
shown in this section that we can estimate the true state of process by 
estimating the perturbation vector, which is governed by a linear differential 
equation. The next task is to obtain a linear perturbation observation from 
the non-linear observation scheme.
\subsection{The observation perturbation vector }
In this section we will adopt the notation $X_{n}$ and $Y_{n}$ for $X(t_{n})$
 and $Y(t_{n})$ respectively.
We define a simulated noise free observation vector $\bar {Y}_{n}$  as follows:

\begin{equation}
\bar {Y}_{n}=G(\bar {X}_{n}) 
\end{equation}

Subtracting $\bar {Y}_{n}$ from the actual observation $Y_{n}$ gives the  
{\em observation perturbation vector}:

\begin{equation}
\delta Y_{n}=Y_{n}-\bar {Y}_{n}
 \end{equation} 

In appendix A we show that the observation perturbation vector is related to the 
state perturbation vector as follows:

\begin{equation}
\delta Y_{n}=M(\bar{X}_{n})\delta X_{n}+v_{n} 
\end{equation} 

where $M(\bar{X}_{n})$ is the Jacobean matrix of G, evaluated at $\bar{X}_{n}$. 
The matrix is also called the  {\em observation sensitivity matrix} and is defined as follows: 

\begin{equation}
M(\bar{X}_{n})=\left.\frac{\partial F(X_{n})}{\partial(X_{n})}\right|_{\bar{X}_{n}}
\end{equation}

We now examine the sequence of observations.
\subsection{Sequence of observation}
We assume that $L+1$ observation are obtained with time stamps 
$t_{n}, t_{n-1},...,t_{n-L}$. Theses observations are assembled as follows :

\begin{equation}\label{eq:obs}
\left[\begin{array}{c}
\delta Y_{n}\\
\delta Y_{n-1}\\
.\\
.\\
.\\
\delta Y_{n-L}\end{array}\right]=\left[\begin{array}{c}
M(\bar{X}_{n})\delta X_{n}\\
M(\bar{X}_{n-1})\delta X_{n-1}\\
.\\
.\\
.\\
M(\bar{X}_{n-L})\delta X_{n-L}\end{array}\right]+\left[\begin{array}{c}
v_{n}\\
v_{n-1}\\
.\\
.\\
.\\
v_{n-L}\end{array}\right]
\end{equation}

Using the relationship: 
\begin{equation}
\delta X_{m}=\Phi (t_{m},t_{n},\bar{X})\delta X_{n}
\end{equation}

then, substituting Equation\ \ref{eq:obs} the observation sensitity equation can be written as:
\begin{equation}
\mathbf{\delta Y}_{n}=\mathbf{T}_{n}\delta X_{n}+\mathbf{V}_{n}
\end{equation}

in which  $\mathbf{T}_{n}$, the {\em total observation matrix} is defined as follows:

\begin{equation}
\mathbf{T}_{n}=\left[\begin{array}{c}
M(\bar{X}_{n})\\
M(\bar{X}_{n-1})\Phi(t_{n-1},t_{n};\bar{X})\\
.\\
.\\
.\\
M(\bar{X}_{n-L})\Phi(t_{n-L},t_{n};\bar{X})\end{array}\right]
\end{equation}

The vectors  $\mathbf{\delta Y}_{n}$ and $\mathbf{V}_{n}$ are large. The solution of the equation can be obtained  from the minimum variance estimation as follows:

\begin{equation}
\delta\hat{X}_{n}=(\mathbf{T}_{n}^{T}\mathbf{R}_{n}^{-1}\mathbf{T}_{n})^{-1}
\mathbf{T}_{n}^{T}\mathbf{R}_{n}^{-1}\mathbf{\delta Y}_{n}
\end{equation}
The estimate $\delta\hat{X}_{n}$ has a covariance matrix:

\begin{equation}
S_{n}=(\mathbf{T}_{n}^{T}\mathbf{R}_{n}^{-1}\mathbf{T}_{n})^{-1}
\end{equation}
where $\mathbf{R}_{n}^{-1}$ is a block diagonal weight matrix, also called the {\em least 
squares weight matrix}, but, in fact, if we define $R_{n}$ as the covariance matrix of 
the the error vector $v_{n}$. Then $\mathbf{R}_{n}^{-1}$  is expressed as:

\begin{equation}
\mathbf{R}_{n}^{-1}=\left[\begin{array}{cccccc}
R_{n}^{-1} & 0 & . & . & . & 0\\
0 & R_{n-1}^{-1} &  &  &  & .\\
. &  & . &  &  & .\\
. &  &  & . &  & .\\
. &  &  &  & .\\
0 & . & . & . & 0 & R_{n-L}^{-1}\end{array}\right]
\end{equation}

In this section we arrived at a form of filter that uses the minimum variance 
estimation initiated by Gauss and the local linearisation technique championed 
by Newton, to estimate the estate of the process from the non linear observation scheme.
This filter is called Gauss-Newton filter (GNF) and is described in detail in Morrison's work \cite{Morrison:newbook, Morrison:1969}. The GNF  has been 
successfully implemented in some practical applications:

\cite{Morrison:2007a}
 showing strong stability. 
The memory nature of the filter has made it unattractive to researchers in the past, and even now, challenging \cite{Milburn:2010a} . However recent developments have presented  recursive form of the linear 
least-squares for state space model \cite{malik:04} . We derive a recursive form of GNF using a similar approach to M. B. Malik \cite{malik:04}. 
However, before we derive a recursive form of the GNF filter, we rewrite the expression of 
$\mathbf{T}_{n}$ using the backward differentiation:

\begin{equation}
\Phi (t_{n-L},t_{n},\bar{X})=A(\bar{X}_{n-L})^{-1}\Phi (t_{n-L+1},t_{n},\bar{X})
\end{equation}

The expression is thus:
\begin{equation}\mathbf{\delta Y}_{n}
\label{T_expression}
\mathbf{T}_{n}=\left[\begin{array}{c}
M_{0}\\
M_{1}A_{1}\\
M_{2}A_{2}\\
.\\
.\\
.\\
M_{L}A_{L}\end{array}\right]
\end{equation}
where

\begin{equation}
A_{L}=\prod_{i=1}^{L}A(\bar{X}_{n-i})^{-1}
\end{equation}

and 

\begin{equation}
M_{L}=M(\bar{X}_{n-L})
\end{equation}
with
\begin{equation}
A_{0}=I
\end{equation}

We now move to derive the {\em Recursive Gauss Newton Filter} in the next section.
\section{The Recursive Gauss-Newton filter\label{sec:recurse}}
 To obtain the recursive  form, we use an approach similar to  M. B. Malik in \cite{malik:04}. Suppose that the
 observations start arriving at $n=0$ and that all initial values of the filter 
are available.  In in order to maintain the filter adaptiveness, a weight matrix function using a fading parameter
$\lambda<1$ is adopted, and is defined as follows:

\begin{equation}
\label{weight_exp}
\mathbf{R}_{n}^{-1}=\left[\begin{array}{cccccc}
R^{-1} & 0 & . & . & . & 0\\
0 & \lambda R^{-1} &  &  &  & .\\
. &  & . &  &  & .\\
. &  &  & . &  & .\\
. &  &  &  & .\\
0 & . & . & . & 0 & \lambda ^{n}R^{-1}\end{array}\right]
\end{equation}

The following, further definitions are adopted:

\begin{equation}
\label{W_expression}
\mathbf{W}_{n}=\mathbf{T}_{n}^{T}\mathbf{R}_{n}^{-1}\mathbf{T}_{n}
\end{equation}

\begin{equation}
\label{zi_expression}
\mathbf{\mathbf{\xi}}_{n}=\mathbf{T}_{n}^{T}\mathbf{R}_{n}^{-1}\delta\mathbf{Y}
\end{equation} 

Resulting in:
\begin{equation}
\label{deltax_expression}
\delta \hat{X}_{n}=\mathbf{W}_{n}^{-1}\xi_{n}
\end{equation}.

In the next section, the recursive update of the perturbation vector is demonstrated.

\subsection{The recursive update of $\mathbf{W}_{n}$}
Using equation (\ref{T_expression}) and the definitions in equations (\ref{W_expression}) and (\ref{weight_exp}) we have:
\setlength {\arraycolsep}{0.0em}

\begin{eqnarray}
\label{W_expansion}
\mathbf{W}_{n}&{}=&{}\sum_{j=1}^{L}\lambda ^{j}R^{-1}\prod_{i=1}^{j}A(\bar{X}_{n-i})^{-T}M(\bar{X}_{n-j})^{T} \nonumber \\
&&\times {}M(\bar{X}_{n-j})\prod_{i=0}^{j}A(\bar{X}_{n-i})^{-1} \nonumber \\
&&{+}\:M(\bar{X}_{n})^{T}R^{-1}M(\bar{X}_{n})
\end{eqnarray}

\setlength {\arraycolsep}{2pt}
and 
\setlength {\arraycolsep}{0.0em}
\begin{eqnarray}
\label{Wminus_expansion}
\mathbf{W}_{n-1}&{}=&{}\sum_{j=1}^{L-1}\lambda ^{j}R^{-1}\prod_{i=1}^{j}A(\bar{X}_{n-1-i})^{-T}M(\bar{X}_{n-1-j})^{T} \nonumber \\
&&\times{}^{-1}M(\bar{X}_{n-1-j})\prod_{i=0}^{j}A(\bar{X}_{n-1-i})^{-1} \nonumber \\
&&{+}\:M(\bar{X}_{n-1})^{T}R^{-1}M(\bar{X}_{n-1})
\end{eqnarray}
\setlength {\arraycolsep}{2pt}

Comparing equations (\ref{W_expansion}) and (\ref{Wminus_expansion}) the following recursive equation is obtained:

\begin{equation}
\label{lyapunov}
\mathbf{W}_{n}=\lambda A(\bar{X}_{n-1})^{-T}\mathbf{W}_{n-1}A(\bar{X}_{n-1})^{-1}+
M(\bar{X}_{n})^{T}R^{-1}M(\bar{X}_{n})
\end{equation}

which is the  discrete, quadratic, Lyapunov, difference equation.

\subsection{The recursive form of $\mathbf{\mathbf{\xi}}_{n}$}
Using equations (\ref{T_expression}) (\ref{weight_exp}) (\ref{zi_expression}) $\mathbf{\mathbf{\xi}}_{n}$ can be expressed as:

\setlength {\arraycolsep}{0.0em}
\begin{eqnarray}
\label{zi_equation}
\xi_{n}&{}=&{}\sum_{j=0}^{L}\lambda ^{j}R^{-1}\prod_{i=1}^{j}A(\bar{X}_{n-i})^{-T}M(\bar{X}_{n-j})^{T}\delta Y_{n-j} \nonumber \\
&&{+}\:M(\bar{X}_{n})^{T}R^{-1}\delta Y_{n} 
\end{eqnarray}
and
\setlength {\arraycolsep}{2pt}
\setlength {\arraycolsep}{0.0em}
\begin{eqnarray}
\label{ziminus_equation}
\xi_{n-1}&{}=&{}\sum_{j=1}^{L-1}\lambda ^{j}R^{-1}\prod_{i=1}^{j}A(\bar{X}_{n-1-i})^{-T}M(\bar{X}_{n-1-j})^{T}\nonumber\\
&&{\times}\: \delta Y_{n-1-j}+M(\bar{X}_{n-1})^{T}R^{-1}\delta Y_{n-1} 
\end{eqnarray}
\setlength {\arraycolsep}{2pt}

Comparing equations (\ref{zi_equation}) and (\ref{ziminus_equation}) the following recursive equation is obtained:

\begin{equation}
\label{recursive_zi}
\xi_{n}=\lambda A(\bar{X}_{n-1})^{-T}\xi_{n-1}+M(\bar{X}_{n})^{T}R^{-1}\delta Y_{n}\end{equation}

\section{Adaptation to Levenberg and Maquardt\label{sec:Maquard}}
In order to guarantee local convergence of the recursive filter and also to avoid the singularity of $\mathbf{W}_{n}$. we replace it by $\mathbf{W}_{n}+\mu I$ as suggested by Levenberg and Maquardt. The presence of the damping factor $\mu$ will have two effects:

\begin{itemize}
\item for large value of $\mu$ the algorithm behaves as a steepest descent which is ideal when the current solution is far from the local minimum. The convergence will be slow but however guaranteed. We therefore have

\begin{equation}
\delta \hat{X}_{n}=\frac{1}{\mu}\xi_{n}
\end{equation}.
\item for $\mu$ very small the algorithm will behave as gauss newton with faster convergence. The current step will be 
\begin{equation}
\delta \hat{X}_{n}=\mathbf{W}_{n}^{-1}\xi_{n}
\end{equation}.
\end{itemize}

\subsection{The Gain Ratio} 
The $\mu$ can be updated by the so called gain ratio. We consider the following cost function which is 

\begin{equation}
E(\delta X_{n})=(\mathbf{\delta Y}_{n}-\mathbf{T}_{n}\delta X_{n})^{T}{R}^{-1}(\mathbf{\delta Y}_{n}-\mathbf{T}_{n}\delta X_{n})\end{equation}

The denominator of gain ratio is :
\begin{equation}
E(0)-E(\delta X_{n})=\delta X_{n}^{T}(\xi_{n}+\mu\delta X_{n})\end{equation}

We define :
\begin{equation}
F(\delta X_{n})=(Y_{n}-G(\bar{X}_{n}+\delta X_{n}))^{T}R^{-1}(Y_{n}-G(\bar{X}_{n}+\delta X_{n}))
\end{equation}

The gain ratio is therefore: 

\begin{equation}
\varrho=\frac{F(0)-F(\delta X_{n})}{E(0)-E(\delta X_{n})}\end{equation}

A large value of $\varrho$ indicates that $E(\delta X_{n})$ is a good approximation of $\bar{Y}$, and $\mu$ can be decreased  so that the next Levenberg-Marquardt step is closer to the Gauss-Newton step. If $\varrho$ is small or negative then $E(\delta X_{n})$ is a poor approximation, then $\mu$ should be increased to move closer to the steepest descent direction. The complete filter algorithm adapted from \cite{LMA}  is presented  in Algorithm  \ref{alg:algorithm1}

\begin{algorithm}
$k:=0$;$\nu:=2$;$\bar{X}_{n}:=X_{n/n-1}$;

$\delta{Y}_{n}:={Y}_{n}-G(\bar{X}_{n})$;

${W}_{temp}=M(\bar{X}_{n})^{T}R^{-1}M(\bar{X}_{n})$;

$\mathbf{W}_{n}=\mathbf{W}_{n-1/n}+{W}_{temp}$;

$\xi_{temp}=M(\bar{X}_{n})^{T}R^{-1}\delta Y_{n}$;

$\xi_{n}=\xi_{n/n-1}+\xi_{temp}$;

$stop:=false$;$\mu=\tau*max(diag(\mathbf{W}_{n/n-1}))$;

While (not stop) and ($k\leq k_{max}$)

~~~~$k:=k+1$;

~~~~repeat;

~~~~solve $(\mathbf{W}_{n}+\mu I)\delta\hat{X}_{n}=\xi_{n}$;

~~~~if ($||\delta\hat{X}_{n}||\leq\varepsilon||\bar{X}_{n}||$)

~~~~~~~~~stop:=true;

~~~~else

~~~~~~~~~~$X_{new}:=\bar{X}_{n}+\delta\hat{X}_{n}$;

~~~~~~~~~~~~~$F(\delta X)=Y_{n}-G(X_{new})$;$F(0)=\delta Y_{n}^{T}R^{-1}\delta Y_{n}$;

~~~~~~~~~~~~~~$E(0)-E(\delta X_{n})=\delta X_{n}^{T}(\xi_{n}+\mu\delta X_{n})$;

~~~~~~~~~~~~~$\varrho=\frac{F(0)-F(\delta X_{n})}{E(0)-E(\delta X)}$;

~~~~~~~~~~~~~if $\varrho>0$

~~~~~~~~~~~~~~~~~~~~~$\bar{X}_{n}=X_{new}$;

~~~~~~~~~~~~~~~~~~~~~~$\delta{Y}_{n}:={Y}_{n}-G(\bar{X}_{n})$;

~~~~~~~~~~~~~~~~~~~~~~${W}_{temp}=M(\bar{X}_{n})^{T}R^{-1}M(\bar{X}_{n})$;

~~~~~~~~~~~~~~~~~~~~~~$\mathbf{W}_{n}=\mathbf{W}_{n/n-1}+{W}_{temp}$;

~~~~~~~~~~~~~~~~~~~~~~$\xi_{temp}=M(\bar{X}_{n})^{T}R^{-1}\delta Y_{n}$;

~~~~~~~~~~~~~~~~~~~~~~$\xi_{n}=\xi_{n/n-1}+\xi_{temp}$;

~~~~~~~~~~~~~~~~~~~~~~$\mu=\mu*max(1/3,1-(2\varrho+1)^{3})$;$\nu:=2$;

~~~~~~~~~~~~~else

~~~~~~~~~~~~~~~~~~~~~~~$\mu:=\nu*\mu$;

~~~~~~~~~~~~~~~~~~~~~~~~$\nu:=2*\nu$;ssm

~~~~~~~~~~~~endif

~~~~~~~endif

~~~~until($\varrho>0$)or(stop);

endwhile

$X_{n/n}=X_{new}$;

$X_{n/n+1}=\Phi(s)X_{n/n}$;

$\mathbf{W}_{n/n+1}=\lambda A({X}_{n/n})^{-T}\mathbf{W}_{n}A({X}_{n/n})^{-1}$;

$\xi_{n/n+1}=\lambda A({X}_{n/n})^{-T}\xi_{n}$;

\caption{L-M algorithm for tracking system}
\label{alg:algorithm1}
\end{algorithm}

\section{State Space Models\label{sec:ssm}} 
We will present four possible models to which the recursive GNF can be applied:

 \begin{itemize}
\item Model 1, with linear process dynamic and linear observation scheme. In this model 
the recursive formulation is similar to the derived forms except the estimation is made directly for $X_{n}$ and that the observed perturbation vector $\delta Y_{n}$ is replaced by the actual observation vector $Y_{n}$.The sensitivity matrices in this case become the measurement and transition matrices of the process. In this case the LM algorithm is not required.
\item Model 2, with linear process dynamic and non-linear observation scheme. The recursive model of the filter remains the same except the state sensitivity matrix becomes a the transition matrix of the process. The state perturbation is estimated to obtain the estimate of the process state.
\item Model 3, with non-linear process dynamic and linear observation scheme. The measurement sensitivity matrix has become 
the measurement matrix.
\item Model 4, with a non-linear process dynamic and non linear observation scheme. 
The derived recursive form without any further modification is applicable to this case. 
\end{itemize}
\subsection{Stability of the Recursive GNF} 
The matrix $\mathbf{W}_{n}$ is the inverse of of the covariance matrix of the filter and is therefore positive definite. 
As a consequence the solution of the derived discrete Ly\ref{sec:ssm}apunov equation in (\ref{lyapunov}) is unique with the sensitivity matrix being stable.
The eigenvalues of the inverse of the  sensitivity matrix are within an open unit circle and therefore the stability of athe system
is ensured by having $\lambda <1$.

\section {Simulation: Range and Bearing tracking\label{sec:demo}}

In these simulation studies, we consider an example of a vehicle  executing various manoeuvres. During turn manoeuvres of unknown constant turn rate,  the aircraft dynamic model is :

\begin{equation}
X_{n}=\left[\begin{array}{ccccc}
1 & \frac{sin(\Omega T)}{\Omega} & 0 & -(\frac{1-cos(\Omega T)}{\Omega}) & 0\\
0 & cos(\Omega T) & 0 & -sin(\Omega T) & 0\\
0 & \frac{1-cos(\Omega T)}{\Omega} & 1 & \frac{sin(\Omega T)}{\Omega} & 0\\
0 & sin(\Omega T) & 0 & cos(\Omega T) & 0\\
0 & 0 & 0 & 0 & 1\end{array}\right]X_{n-1}+v_{n}\end{equation}

where the state of the vehicle is $X_{n}=[x,\dot{x},y,\dot{y},\Omega]$, with $x$,$y$ the position coordinates and $\dot{x}$,$\dot{y}$ their corresponding velocity components.The process noise $v_{k}\sim\mathcal{N}(0,Q)$ with covariance matrix $Q=diag\left[\begin{array}{ccc}
q1BB^{T} & q1BB^{T} & q2T\end{array}\right]$  where,

\begin{equation}
BB^{T}=\left[\begin{array}{cc}
\frac{T^{4}}{4} & \frac{T^{3}}{2}\\
\frac{T^{3}}{2} & T^{2}\end{array}\right]
\end{equation}

When the vehicle moves at nearly constant velocity its dynamic model is:
\begin{equation}
X_{n}=\left[\begin{array}{ccccc}
1 & T & 0 & 0 & 0\\
0 & 1 & 0 & 0 & 0\\
0 & 0 & 1 & T & 0\\
0 & 0 & 0 & 1 & 0\\
0 & 0 & 0 & 0 & 1\end{array}\right]X_{n-1}+v_{n}
\end{equation}

The vehicle  is observed by a radar located at the origin of the plane, capable of measuring the range $r$ and and the bearing angle $\theta$. The measurement equation is therefore: 

\begin{equation}
\left[\begin{array}{c}
r_{n}\\
\theta_{n}\end{array}\right]=\left[\begin{array}{c}
\sqrt{x^{2}+y^{2}}\\
tan^{-1}(\frac{y}{x})\end{array}\right]+w_{n}
\end{equation}

where the measurement noise is $w_{k}\sim\mathcal{N}(0,R)$ with covariance 

$R=diag\left[\begin{array}{cc}

\sigma_{r}^{2} & \sigma_{\theta}^{2}\end{array}\right]$

The following constants were used  for data generation:
$T=1s$; $\Omega=-3^{0}s^{-1}$; $q1=0.01$m$^{2}\rm{s}^{-4}$; $q2=1.75\times10^{-4}\rm{s}^{-4}$; $\sigma_{r}$=10\rm{m}; $\sigma_{\theta}=\sqrt{0.1}$mrad.

The vehicle starts at true initial state  $X_{n}$=[10m, 25ms$^{-1}$,400m, 0ms$^{-1}$,-3ms$^{-1}$] and moves at nearly constant velocity for $100$s, Then it executes a turn manoeuvre from time index $n=101$ to $n=150$. After the manoeuvre,  the vehicle's velocity remains nearly constant from  $n=151$ to  $n=250$. At $n=251$ it starts a new turn manoeuvre  at rate $\Omega$=3\rm{m}s$^{-1} $ until $n=400$. Finally from $n=400$ to  $n=500$ it moves at nearly constant velocity. Figure [1] describes the complete trajectory of the vehicle.

The filter uses a single model of a constant velocity to track the entire manoeuvre:
\begin{equation}A(X_{n})=\left[\begin{array}{ccccc}
1 & T & 0 & 0 & 0\\
0 & 1 & 0 & 0 & 0\\
0 & 0 & 1 & T & 0\\
0 & 0 & 0 & 1 & 0\\
0 & 0 & 0 & 0 & 1\end{array}\right]\end{equation}
The initial value $\mathbf{W}_{-1/0}=10^{-2}I$, where $I$ is an identity matrix.
The filter parameters are the following $k_{max}=200$, $\varepsilon=1\times10^{-24}$, $\tau=1\times10^{-3}$, $\lambda=0.4$
The filter initial state is generated randomly and then ensuring that it has the same sign as the true state. This procedure guarantees the local convergence of the first estimate.
The experiment was repeated for 250 Monte Carlo runs and  the root means squared error (RMSE) is used as a performance metric. The position RMSE is computed using the following expression:
\begin{equation}
RMSE=\sqrt{\frac{1}{N}\sum_{i=1}^{N}\left((x_{n}^{i}-\hat{x_{n}})^{2}+(y_{n}^{i}-\hat{y_{n}})^{2}\right)}\end{equation}
where $(x_{n}^{i},y_{n}^{i})$ and $(\hat{x_{n}},\hat{y_{n}})$ true and estimated position coordinates respectively.
The velocity root mean square error (RMSE) is computed similarly.
Figures [2] and [3] show the RMSE of the position and velocity respectively. The position RMSE is not affected by different manoeuvres while the velocity RMSE shows variation from different manoeuvre states. The average values of the damping factor  after  complete cycles of iteration is presented in Figure [3]. The damping factor increases rapidly at the transition between manoeuvres. The average number of iterations $k$ at convergence from Figure [4] shows similar variations.

\section{Conclusions}
The GNF with memory  combines the minimum 
variance estimation and  the Newton method of local linearisation to estimate the process
true state. The recursive form for the Gauss-Newton filter has been derived in one 
compact form that can be applied to all the four state and observation linearity
and nonlinearity scenarios:

\begin{description}
\item Case 1 : linear process dynamic and linear observation scheme. 
\item Case 2 : linear process dynamic and non-linear observation scheme. 
\item Case 3 : with non-linear process dynamic and linear observation scheme. 
\item Case 4 : non-linear process dynamic and non linear observation scheme. 
\end{description}

The Hessian matrix of the filter which is computed recursively is augmented by a damping factor as suggested earlier by Levenberg-Maquardt for non linear curve fitting problems. The new filter is therefore a combination of Newtons steepest descent and  the Gauss-newton, ensuring its robustness. The presence of a forgetting factor in the filter equations renders it capable of tracking manoeuvring targets  with a single filter dynamic model.


%

\appendix
\section{}
\subsection {The differential equation governing $\delta X(t)$ \label{ap:partA}}
Starting from:
\begin{equation}
\delta X(t)=X(t)-\bar{X}(t)\end{equation}
The differentiation rule is applied:

\begin{equation}
D\delta X(t)=F(\bar{X}(t)+\delta X(t))-F(\bar{X}(t))\end{equation}
Let $F$ be defined as follows :

\begin{equation}
F=\left[\begin{array}{c}
f_{1}\\
.\\
.\\
.\\
f_{n}\end{array}\right]\end{equation}
Equation becomes:
\begin{equation}
D\delta X(t)=\left[\begin{array}{c}
f_{1}(\bar{X}(t)+\delta X(t))\\
.\\
.\\
.\\
f_{n}(\bar{X}(t)+\delta X(t))\end{array}\right]-\left[\begin{array}{c}
f_{1}(\bar{X}(t))\\
.\\
.\\
.\\
f_{n}(\bar{X}(t))\end{array}\right]\end{equation}
The Taylor first order approximation is applied:
\setlength {\arraycolsep}{0.0em}
\begin{eqnarray}
D\delta X(t)&{}=&{}\left[\begin{array}{c}
f_{1}(\bar{X}(t))\\
.\\
.\\
.\\
f_{n}(\bar{X}(t))\end{array}\right]+\left[\begin{array}{c}
\nabla f_{1}(\bar{X}(t))^{T}\\
.\\
.\\
.\\
\nabla f_{n}(\bar{X}(t))^{T}\end{array}\right]\delta X(t) \nonumber \\
&&{-}\:\left[\begin{array}{c}
f_{1}(\bar{X}(t))\\
.\\
.\\
.\\
f_{n}(\bar{X}(t))\end{array}\right]\end{eqnarray}
\setlength {\arraycolsep}{5pt}
The following relation is obtained :
\begin{equation}
D\delta X(t)=A(\bar{X}(t))\delta X(t)\end{equation}
Where:
\begin{equation}
A(\bar{X}(t))=\left[\begin{array}{c}
\nabla f_{1}(\bar{X}(t))^{T}\\
.\\
.\\
.\\
\nabla f_{n}(\bar{X}(t))^{T}\end{array}\right]=\left.\frac{\partial F(X(t))}{\partial(X(t))}\right|_{\bar{X}(t)}\end{equation}

\subsection {The relation between  $\delta X_{n}$ and $\delta Y_{n}$}
\begin{equation}
\delta Y_{n}=G(\bar{X}_{n}+\delta X_{n})-G(\bar{X}_{n})\end{equation}

As direct consequence of \ref{ap:partA} the following relationship is obtained:

\begin{equation}
\delta Y_{n}=M(\bar{X}_{n})\delta X_{n}+v_{n}\end{equation}

\section{Figure captions list}

Figure 1: Target complete trajectory with manoeuvres

Figure 2: The Position RMSE is unaffected by the manoeuvres.

Figure 3: The velocity RMSE varies with manoeuvres.

Figure 4: The damping factor shows sharp peaks at start of manoeuvres.

Figure 5: The number of iterations increases during manoeuvres.

The figure numbering  appears in the same order as the figures in the  pdf document


\section*{Acknowledgment}

The authors would like to thank Dr Norman Morrison for his contribution during
the research that leads to obtaining a recursive form of GNF. 
Dr Morrison has been  working on the GNF throughout his career and even in his retirement
 is enthusiastic in providing teaching and  insights into  the fundamentals of filter  Engineering.

\bibliographystyle{elsarticle-num}
\bibliography{Reference}

\begin{thebibliography}{10}
\expandafter\ifx\csname url\endcsname\relax
  \def\url#1{\texttt{#1}}\fi
\expandafter\ifx\csname urlprefix\endcsname\relax\def\urlprefix{URL }\fi
\expandafter\ifx\csname href\endcsname\relax
  \def\href#1#2{#2} \def\path#1{#1}\fi

\bibitem{Yunmin:99}
{Recursive Least Squares With Linear Constraints}.

\bibitem{Thomas:00}
T.~Kailath, A.~H. Sayed, B.~Hassibi, Linear Estimation, 2nd Edition, Prentice
  Hall, New Jersey, USA, 2000.

\bibitem{malik:04}
M.~B. Malik, State-space recursive least-squares: {P}art i, Signal Processing
  84 (2004) 1709--1718.

\bibitem{Morrison:newbook}
N.~Morrison, Filter engineering -a practical approach: The {G}auss-{N}ewton and
  polynomial filters, to be published.

\bibitem{Morrison:2007a}
N.~Morrison, R.~T. Lord, M.~R. Inggs, The {G}auss-{N}ewton algorithm applied to
  track-while-scan radar, in: Proceedings of the IET International Conference
  on Radar Systems (RADAR 2007), Institution for Engineering and Technology,
  2007.

\bibitem{jpdacon:11}
J.-P. da~Conceicao, Accelerating {G}auss-{N}ewton filters on {FPGAs}, Master's
  thesis, University of Cape Town (Dec. 2011).

\bibitem{Milburn:2010a}
J.~Milburn, Co-processor offloading applied to passive coherent location with
  doppler and bearing data, Master's thesis, University of Cape Town - RRSG
  (February 2010).

\bibitem{Mquardt}
D.~W. Marquardt, An algorithm for least-squares estimation of nonlinear
  parameters, Journal of the Society for Industrial and Applied Mathematics
  11~(2) (1963) pp. 431--441.

\bibitem{LMA}
O.~T. K.~Madsen, H.B.~Nielsen, Method of non-linear least squares problems, 2nd
  Edition, Informatics and Mathematical Modelling, Technical University of
  Denmark, 2004.

\bibitem{ScienceDLMapplications}
B.~G. Kermani, S.~S. Schiffman, H.~T. Nagle,
  \href{http://www.sciencedirect.com/science/article/pii/S0925400505000961}{Pe%
rformance of the {L}evenberg-{M}arquardt neural network training method in
  electronic nose applications}, Sensors and Actuators B: Chemical 110~(1)
  (2005) 13 -- 22.
\newblock \href {http://dx.doi.org/10.1016/j.snb.2005.01.008}
  {\path{doi:10.1016/j.snb.2005.01.008}}.
\newline\urlprefix\url{http://www.sciencedirect.com/science/article/pii/S09254%
00505000961}

\bibitem{ScienceDLMapplications2}
E.~Derya, Übeyli,
  \href{http://www.sciencedirect.com/science/article/pii/S1051200408001243}{An%
alysis of {EEG} signals by implementing eigenvector methods/recurrent neural
  networks}, Digital Signal Processing 19~(1) (2009) 134 -- 143.
\newblock \href {http://dx.doi.org/10.1016/j.dsp.2008.07.007}
  {\path{doi:10.1016/j.dsp.2008.07.007}}.
\newline\urlprefix\url{http://www.sciencedirect.com/science/article/pii/S10512%
00408001243}

\bibitem{ScienceDLMapplications3}
V.~Singh, I.~Gupta, H.~Gupta,
  \href{http://www.sciencedirect.com/science/article/pii/S095219760600114X}{{A%
NN}-based estimator for distillation using {L}evenberg-{M}arquardt approach},
  Engineering Applications of Artificial Intelligence 20~(2) (2007) 249 -- 259.
\newblock \href {http://dx.doi.org/10.1016/j.engappai.2006.06.017}
  {\path{doi:10.1016/j.engappai.2006.06.017}}.
\newline\urlprefix\url{http://www.sciencedirect.com/science/article/pii/S09521%
9760600114X}

\bibitem{Morrison:1969}
N.~Morrison, Introduction to Sequential Smoothing and Prediction, McGraw-Hill
  Book Company, 1969.

\end{thebibliography}

\end{document}